\documentclass[12pt]{article}

\usepackage{latexsym}
\usepackage{amsfonts}
\usepackage{amssymb}
\usepackage{verbatim}
\usepackage{cite}
\author{Jonathan P. Hsu$^1$\footnote{pihsu at stanford.edu}, Alexander
Maloney$^{1,2}$\footnote{maloney at slac.stanford.edu} and Alessandro
Tomasiello$^{1}$\footnote{tomasiel at stanford.edu}}
\title{\textbf{Black Hole Attractors and Pure Spinors}}

\def\beqn{\begin{equation}}
\def\eeqn{\end{equation}}

\def\calN{\mathcal{N}}

\def\calP{\mathcal{P}}

\def\Zbar{\bar{Z}}

\def\Cbar{\bar{C}}
\def\Omegabar{\bar{\Omega}}
\def\Re{\mathrm{Re} \,}
\def\Im{\mathrm{Im} \,}

\def\half{\frac{1}{2}}
\def\fourth{\frac{1}{4}}
\def\slash#1{\rlap
{\begin{picture}(10,10)
\put(0,0){\line(1,1){10}} 
\end{picture}} #1}

\def\nn{{\cal N}} 
\def\rr {{\Bbb R}}

\def\del {\partial} 
\def\cy {Calabi--Yau} 
\def\ka {K\"ahler}
\def\vol {\mathrm{vol}}

\def\sei{e^{i J}\!\!\!\!\! 
\begin{picture}(10,10)
\put(0,0){\line(1,2){5}}
\end{picture}
}

\def\prod#1#2{{\left\langle#1,#2\right\rangle}}
\def\O{\Omega}
\def\bO{{\bar \Omega}}
\def\p{\partial}
\def\a{\alpha}
\def\ba{{\bar a}}
\def\bb{{\bar b}}
\def\bt{{\bar t}}

\def\barf{{\bar f}}
\def\bZ{{\bar Z}}
\def\bchi{{\bar \chi}}

\makeatletter
\@addtoreset{equation}{section}
\makeatother

\begin{document}
{\hfill SU-ITP-06/04}

{\hfill SLAC-PUB-11678}
\linebreak[4]
\linebreak[4]
\linebreak[4]
\begin{center}
{\Large{\bf{Black Hole Attractors and Pure Spinors}}}
\linebreak[4]
\linebreak[4]
\linebreak[4]
Jonathan P. Hsu$^{\dagger}$\footnote{pihsu at stanford.edu}, Alexander
Maloney$^{\dagger*}$\footnote{maloney at slac.stanford.edu} and Alessandro
Tomasiello$^{\dagger}$\footnote{tomasiel at stanford.edu}

$^{\dagger}$ \textit{ITP, Stanford University, Stanford, CA 94305, USA}

$^*$ \textit{SLAC, 2575 Sand Hill Rd., Menlo Park, CA 94025}
\linebreak[4]
\end{center}

\begin{abstract}
We construct black hole attractor solutions for a wide class of $\nn=2$ 
compactifications. The analysis is carried out in ten dimensions and makes crucial 
use of pure spinor techniques.  This formalism can accommodate
non-\ka\ manifolds as well as compactifications with flux, in addition to
the usual \cy\ case. 
At the attractor point, the charges fix the moduli according to
$\sum f_k = \Im(C \Phi)$, where $\Phi$ is a pure spinor of odd (even) chirality
in IIB (A). For IIB on a Calabi--Yau, $\Phi=\Omega$ and the equation reduces to the
usual one.  
Methods in generalized complex geometry can be used to study solutions to 
the attractor equation.
\end{abstract}

\newpage
\section{Introduction}

The attractor mechanism is a general feature of black
hole solutions to four dimensional $\calN=2$ supergravity
\cite{att1,att2,att3}.
It states that near the horizon of a supersymmetric black hole the 
vector multiplet moduli flow to special values which only depend
on the charge of the black hole and not on the asymptotic values of 
the moduli.  
The simplest application of the attractor mechanism is to
compactifications of type II string theory on a Calabi-Yau manifold $Y$.
In this case the ten dimensional action of string theory 
reduces, in the low energy limit, to an effective 
$\calN=2$ supergravity theory in four dimensions,
whose field content and action depend on the choice of $Y$.
The attractor mechanism has an elegant interpretation in terms 
of the geometry of $Y$: for type IIA (IIB), the
K\"ahler (complex) structure of $Y$ flows to an attractor fixed point
at the horizon. The attractor mechanism has also been shown to occur for
some non-supersymmetric but extremal black holes
\cite{nonsusy1,nonsusy2,nonsusy3}.

In this paper we will study supersymmetric black hole attractors
for a broader class of compactifications which preserve $\calN=2$
supersymmetry but are not necessarily Calabi-Yau.  
This class includes both non-K\"ahler compactifications as well as 
compactifications with non-trivial background flux.
Examples of such $\calN=2$ compactifications have been constructed using
T-duality \cite{kst}. Some geometrically more interesting 
non--K\"ahler vacua have also been provided recently in \cite{ckt}, but
they involve $g_s\neq 0$ and hence cannot be used as supergravity solutions.
Although from the four dimensional perspective the resulting 
black hole solutions are exactly as in \cite{att1,att2,att3}, 
the geometric description is less clear than in the Calabi-Yau case.
For example, there is no general description of the vector multiplet moduli 
space of these compactifications 
in terms of geometric quantities.

For this reason, we will study these configurations as 
solutions to the full ten dimensional equations of motion, rather than
the low energy effective theory in four dimensions.
From the ten dimensional point of view, these black holes are simply
special classes of solutions with flux, 
to which we can apply the pure spinor techniques of \cite{gmpt}. 
For example, the near horizon geometry of a BPS black hole is just a 
 particular flux compactification whose four dimensional
geometry is $AdS_2\times S^2$.\footnote{
This fact has led to a fruitful interplay between
between the study of flux compactifications and extremal black holes; 
see e.g. \cite{moorelectures, ovv, newattractors,dallagata}.}

The ten dimensional gravitino variations yield a new form of the 
attractor equation, phrased in the language of pure spinors.  
These pure spinors play a central role in the theory of generalized
complex manifolds \cite{hitchin,gualtieri,witt}, and have recently
found several applications in supergravity, in the study of compactifications 
on six \cite{gmpt2,gmpt,jw} and seven \cite{jw,jw2} dimensional manifolds. 
We give a brief introduction to pure spinors in appendix \ref{jazz}.
For practical purposes, a pure spinor $\Phi$ 
may be thought of as formal sum of differential forms of different rank. 

To describe $\nn=2$ compactifications in terms of pure spinors, 
we will follow the approach of \cite{gmpt2}.  These authors classified
$\nn=1$ vacua using a pair of pure spinors $\Phi_\pm$, 
which determine the metric on the internal manifold. 
For type II string theory on a Calabi-Yau, these two pure spinors 
have a simple interpretation. One of them is the
holomorphic three form $\Omega$, which fixes the complex structure of the 
Calabi-Yau, and the other is $e^{iJ}$
where $J$ is the K\"ahler form. 
In general, an $\nn=2$ vacua is characterized by two pairs of pure spinors, 
along with the 
constraint that each pair determines the same metric on the internal manifold.
The BPS black holes under consideration break the $\nn=2$ supersymmetry of a
background down to $\nn=1$.  

The attractor equations describe how the geometry of the 
internal manifold changes as a function of radius.
At every value of $r$, the internal manifold satisfies the 
equations for an $\nn=2$ vacuum in four dimensions.  
However, one linear combination $\Phi$ of the pure 
spinors flows as a function of radius.  So as $r$ changes, the internal manifold flows
through the moduli space of $\nn=2$ compactifications.
At the horizon, this pure spinor approaches a
fixed value determined only by the charge of 
the black hole -- it obeys an equation of the form
\[
\sum_k f_k = \Im (\bar C\Phi)\ 
\]
where $f_k$ is a $k$-form flux.
This equation can be solved in simple geometric
terms, using a theorem of Hitchin \cite{hitchin}.  (Since $\Phi$ is
related to pure spinors describing the vacua, it also obeys an extra 
 differential 
condition, whose general solution is more complicated, as we will see.)
This theorem involves the construction of a function, whose integral --
known as the Hitchin functional --
can be interpreted as the entropy of the associated black hole.\footnote{
This relation between the Hitchin functional and the black hole entropy 
has been noted by \cite{dgnv, pestun1}. 
The Hitchin functional has also found use in other closely related 
physical contexts, see e.g. \cite{pestun2, glw, dgnv}.}
Our construction may be thought of as a physical implementation of this 
theorem;
the attractor equations admit a solutions precisely when the associated 
black hole has a finite area horizon.

The approach described above has 
several advantages, which are relevant even for
standard Calabi-Yau compactifications.
First, because we have solved the full ten dimensional equations
of motion, the solutions apply in cases where the 
four dimensional supergravity equations are no longer valid.
In particular, they can describe configurations where the 
Kaluza-Klein length scale of the compactification manifold
is not small compared to the length scales of the 
four dimensional solution. 
It may therefore prove useful in the study of small
black holes, where the radius of curvature of the black horizon
can be of order the Kaluza-Klein scale (see, e.g.
\cite{Dabholkar:2004yr,Dabholkar:2004dq, Sen:2004dp, Hubeny:2004ji,
Dabholkar:2005by}).  
In addition, this derivation demonstrates explicitly
that BPS black hole solutions can be consistently lifted 
to solutions of the full ten dimensional supergravity.

Our hope is that the universal attractor behavior 
described in this paper may play a role in a better understanding of the 
dynamics and definition of string theory in these backgrounds.
Recently, it has been proposed that such black holes provide a 
non--perturbative definition of topological string theory in the 
Calabi-Yau case \cite{osv}.
It is therefore natural to expect that
the black hole attractors described in this paper are related to topological
string theory on non-Calabi-Yau compactifications.\footnote{
A recent paper \cite{pestun1} has discussed a generalization of 
the conjecture of \cite{osv} in this context, although in absence of RR fields; see also \cite{pestun2}.}

This paper is organized as follows.  
In the next section we will describe the attractor equations in terms 
of pure spinors, and discuss the general properties of these
solutions.
In section 3 we will consider a few simple examples.
Appendix A describes our spinor conventions, and Appendix B contains a brief
introduction to the pure spinor constructions used in the text.
Appendix C reviews a few features of the four dimensional 
attractor equations which are necessary to make contact with the
pure spinor formulation. 

\section{Attractor Black Holes in Ten Dimensions}

In this section we will derive the attractor equations for a wide class of BPS
black holes, using ten dimensional supergravity.  
These equations describe the radial flow of a pure
spinor on the internal manifold.  The derivation 
given below requires some technical manipulations, 
but the main results are rather simple to state.  
For each of the backgrounds under consideration, one can construct eight
pure spinors, which we will call $\Phi^{13}_\pm$, $\Phi^{24}_\pm$,
$\Phi^{14}_\pm$ and $\Phi^{23}_\pm$.  These pure spinors are constructed from
the supersymmetry variations.  The first two of these 
pure spinors obey the constraints required for a compactification to an
$\nn=2$ Minkowski vacuum.  The other two obey a first order differential
equation, which describes how the internal 
geometry flows in the moduli space of $\nn=2$ vacua as a function of radius.
These equations are the attractor equations for this background; from the 
four dimensional point of view, they describe the radial flow of the 
vector multiplet moduli. The explicit 
equations describing this flow are written 
down at the end of section 2.3.

In section 2.1 we describe the basic form of the backgrounds under 
consideration, in section 2.2 we write down the fermion variations, 
and in section 2.3 we rewrite the BPS conditions in terms of pure spinors.
Section 2.4 contains a brief discussion of the solutions of these equations, 
using a theorem of Hitchin's.

\subsection{The Background}

We will start by describing the background under consideration.

We are interested in BPS solutions of type II supergravity
that describe a four dimensional black hole geometry times an internal
six-manifold $Y$.  The ten dimensional metric will be of the form
\beqn
ds^2= e^{2B(y)}\left(-e^{2U(r)} dt^2 + e^{-2U(r)} (dr^2 +
r^2(d\theta^2 + \cos \theta^2 d\phi^2))\right) + g_{mn}(r,y)dy^m dy^n .
\eeqn
The $(t,r,\theta,\phi)$ components of the metric
describe an extremal black hole solution in four dimensions, whose geometry 
depends on the function $U(r)$.
The metric $g_{mn} dy^m dy^n$ on $Y$ is a function of radius
as well as the internal coordinates,  
and we have explicitly included a warp factor $B(y)$.  
Although in principal we could dimensionally reduce on $Y$ to obtain an
effective supergravity in $D=4$,
it turns out to be much easier to study these black hole solutions 
by working directly with ten dimensional quantities.

The spin--connection following from this metric 
has the form 
$D_M = \partial_M + \fourth \Omega_M^{AB}\Gamma_{AB}$, where
$M$ is a curved 10-dimensional index and $A,B$ are flat indices.
The components of the spin connection are\footnote{
Here ${}'$ denotes derivative with respect to $r$. }
\beqn
\begin{array}{c}\vspace{.2cm}
\Omega_t{}^{01} =  e^{2U} U', \quad \Omega_{\theta}{}^{12}= -1 +
r U',\quad
\Omega_{\phi}{}^{13} = \cos\theta (-1 + r U'),\quad
\Omega_{\phi}{}^{23} = \sin\theta \\\vspace{.2cm}
\Omega_{r}{}^{ab} = e^{m [a} 
e_{m}^{b]}{}' = 0,\quad \Omega_{m}{}^{1a} = -\half e^{-B+U} e^{na}
g_{nm}',\quad \Omega_{m}{}^{ab} = \omega_{m}{}^{ab}\\
\Omega_t^{0a}= e^{B+U} e^{am}\del_m B \ , \quad 
\Omega_r^{1a}= e^{B-U} e^{am } \del_m B\ , \quad
\Omega_{\theta}^{2a}= r e^{B-U} e^{am} \del_m B \ , \quad
\Omega_{\phi}^{3a}= r e^{B-U} \cos\theta e^{am} \del_m B \  .
\end{array}
\eeqn
where $m, n$ are curved indices on $Y$, and $a,b$ the associated flat
indices.  The 6-bein $e_a^m$ on $Y$ obeys $e_a^m e_b^n g_{mn}=\delta_{ab}$.
We have chosen our local frame to obey 
$(e_a{}^m)' = \beta^m{}_n e_a{}^n$, where $\beta_{mn} = -\frac12 g_{mn}'$
is symmetric in $mn$.  This is why $\Omega_r{}^{ab}=0$.


In addition to the metric described above, 
the backgrounds under consideration will include flux.
The R-R fluxes can be decomposed as
\beqn\label{eq:fis}
F^{(10)}_{2n} = \vol_A \wedge f^A_{2n-2} + \vol_S \wedge f^S_{2n-2} + F^i_{2n} +
\vol_A\wedge \vol_S \wedge F^e_{2n-4}
\eeqn
where $\vol_A = (e^{2U}/r^2)dt\wedge dr$ and $\vol_S = \cos \theta
d\theta\wedge d\phi$.  Here $f^{A}$, $f^S$, $F^i $ and $F^e$ are differential
forms on $Y$.  A subscript on a form indicates its rank; in the discussion 
below we will often drop these subscripts when they are not necessary. 
The first two terms in (\ref{eq:fis}) are the gauge field produced by
the charged black hole; if we were to dimensionally reduce to 
four dimensions, they would describe electric and magnetic fluxes
sourced by a configuration of branes wrapped on $Y$. The last two terms
describe purely internal and external $2n$-form flux. In type
IIA, the index $n$ runs over $0, 2, 4, 6, 8, 10$ 
while in type IIB $n$ runs over 
$1/2, 3/2, 5/2, 7/2, 9/2$. 
The R-R fluxes described above contain both field strengths and their duals, 
so we must impose the self-duality relations\footnote{
$Int[n]$ denotes the integer part of $n$ and $*_{10}$ the ten dimensional Hodge
star.}
\beqn\label{eq:selfduality}
F^{(10)}_{2n} = (-1)^{Int[n]}*_{10}F_{10-2n} \ .
\eeqn
This relates $f^A$ to $f^S$ and $F^i$ to $F^e$, so from now on we will
write our expressions involving R-R fluxes in terms of $F\equiv F^i$ and
$f\equiv f^S$. 

We will also consider NS-NS fluxes of the form 
\beqn
H^{(10)}= 
H_3 + dr \wedge b_2'
\eeqn 
where $H$ and $b$ are differential forms on $Y$.  
The second term in this expression arises because we are allowing
the internal NS-NS two form $b$ to depend on $r$.
 
The Bianchi identities and source-free 
equations of motion for the R-R fields take the form
$(d-H^{(10)}\wedge)F^{(10)}=0$.  For the fluxes described above, this is
\begin{eqnarray}
  \label{eq:bianchi}
\nonumber  
(d-H\wedge)F
=0\ , &  \quad (d+H\wedge)(e^{4B} *F
)=0\ , & \quad dH=0\ ; \\
 \del_r(e^{-b\wedge}F)=0\ ,& \quad d(e^{2B}*b)=0\ , & \quad d(e^{4B} *H)=
\frac{e^{2(U+B)}}{r^2} \del_r(r^2*b')\ , \\
\nonumber\del_r(e^{-b\wedge}f)=0\ , &\quad   (d-H\wedge)f=0\ , &\quad 
  (d+H\wedge)*f=0\ .
\end{eqnarray}
Here $*$ is the hodge star on $Y$, and we are omitting the $n$ indices
used above.
These identities, together with BPS equations written below, imply the
full ten dimensional equations of motion.

\subsection{The supersymmetry variations}

The gravitino and dilatino  
variations in ten-dimensional type II supergravity are
\beqn
\begin{array}{rcl}
\delta\psi_M &=& (D_M+\fourth H_M \calP)\epsilon +
\frac{e^{\phi}}{16} \sum_n \slash{F}_{2n} \Gamma_M\calP_n{}\epsilon \\
\delta\lambda & = & (\slash{\partial}\phi + \half \slash{H}\calP)\epsilon
+
 \frac{e^{\phi}}{16} \sum_n \Gamma^M\slash{F}_{2n}\Gamma_M \calP_n{}\epsilon .
\end{array}
\eeqn
We have not written the spinor indices explicitly.
Our gamma matrix conventions are described in Appendix A.
We have also suppressed the doublet indices $i=1,2$ on 
the gravitino $\psi_M$, dilatino $\lambda$, 
and supersymmetry parameter $\epsilon$.
For example, $\epsilon=(\epsilon^1,\epsilon^2)$ is a doublet of
ten-dimensional Majorana-Weyl spinors.
The $\calP$ matrices act on these doublet indices, as 
$\calP = \Gamma_{11}$ and $\calP_n = \Gamma_{11}^n\sigma^1$ in type IIA, 
and as $\calP = -\sigma^3$, $\calP_{n} = \sigma^1$ for $(n+1/2)$ even and
$\calP_n = i\sigma^2$ for $(n+1/2)$ odd in type IIB. 


Using the self-duality relation (\ref{eq:selfduality}), 
and putting in the doublet indices explicitly, the 
the gravitino equation can be written as
\beqn
\begin{array}{rcl}
\delta\psi^1_M &=& (D_M\pm\fourth H_M )\epsilon^1 \mp
\frac{e^{\phi}}{8} \Gamma^{01}\frac{e^{2U}}{r^2}\slash{f}\Gamma_M\epsilon^2
+ \slash{F}\Gamma_M\epsilon^2 \\
\delta\psi^2_M &=& (D_M\mp\fourth H_M )\epsilon^2 +
\frac{e^{\phi}}{8} \Gamma^{01}\frac{e^{2U}} {r^2}\slash{f}^{\dagger}\Gamma_M\epsilon^1
\pm \slash{F}^{\dagger}\Gamma_M\epsilon^2 .
\end{array}
\eeqn
The upper sign is for type IIA and the lower sign for IIB.
We have defined
\beqn
\begin{array}{rcl}
\slash{f} &=& \slash{f}^A_0 - \slash{f}^A_2 + \slash{f}^A_4 -
\slash{f}^A_6 \\
\slash{F} &= &\frac{e^{\phi}}{8}\left(\slash{F}^i_0 - \slash{F}^i_2 +
\slash{F}^i_4 - \slash{F}^i_6 \right).
\end{array}
\eeqn
for type IIA, and 
\beqn
\begin{array}{rcl}
\slash{f} &=& \slash{f}^A_1 + \slash{f}^A_3 + \slash{f}^A_5 \\
\slash{F} &= &\frac{e^{\phi}}{8}\left(\slash{F}^i_1 + \slash{F}^i_3 +
\slash{F}^i_5 \right)
\end{array}
\eeqn
for type IIB.  
In the IIB case, $\slash{f_3}$ is anti-hermitian while $\slash{f}_{1,5}$
are hermitian. In IIA, $\slash{f}_{0,4}$ are hermitian while
$\slash{f}_{2,6}$ are anti-hermitian.

Using the expression for the spin connection, we can write out the components
of the gravitino variations in their full glory.  For example, 
\beqn
\begin{array}{rcl}
\delta\psi^1_t& = & e^{2U}\Gamma^{01}\left(-\frac12 U'\epsilon^1 \mp A(r)
\slash{f}
\Gamma_0\epsilon^2\right) + \Gamma_t(\pm\slash{F}\epsilon^2 + \half
\slash\partial B  \epsilon^1) \\
\delta\psi^1_r&=&\partial_r\epsilon^1 \pm A(r) \slash{f}\Gamma_0\epsilon^2
 + \Gamma_r(\pm\slash{F}\epsilon^2 + \half
\slash\partial B \epsilon^1) 
\pm\frac14 \slash b' \epsilon^1
\\
\delta\psi^1_m&=&(D_m \pm\fourth H_m)\epsilon^1 + \slash{F} \Gamma_m \epsilon^2
+ \Gamma^r\left( \frac14\Gamma^n (-g \pm b)_{mn}'\epsilon^1 
\pm A(r)\Gamma^{0}\slash{f}\Gamma_m\epsilon^2\right),
\end{array}
\eeqn
where $A(r) = e^{B +U+\phi}/8r^2$. The expressions for
$\delta\psi_M^2$ are identical, but with $\slash{f} \to
\mp\slash{f}^{\dagger}$, $\slash{F} \to \pm\slash{F}^{\dagger}$ and $H \to
-H$. The angular components of $\delta \psi$ are similar, so we will not
write them down explicitly. 

In a supersymmetric background these fermion
variations vanish.  We are looking for solutions that preserve
half of the four dimensional supersymmetry, 
so only one linear combination of the supersymmetry parameters
$\epsilon^1$ and $\epsilon^2$ will be preserved by the background.
It turns out that the correct linear relation between  
$\epsilon^1$ and $\epsilon^2$ includes the action of
$\Gamma^0$, but not the action of any other four--dimensional gamma
matrices.\footnote{
This is the standard situation for branes 
in $\rr^4$ that extend in time but not in any other spatial directions.} 
This implies that the terms in the gravitino variation containing $\Gamma^1$
must vanish separately from those that do not.  So the 
$\delta\psi_t=0$ and $\delta\psi_m=0$ equations become
\beqn\label{eq:rdependent}
\begin{array}{rcl}
0&=& -\frac12 U'\epsilon^1 \mp A(r) \slash{f} \Gamma_0\epsilon^2 \\
0&=& \frac{1}{4}\Gamma^n (-g\pm b)_{mn}'\epsilon^1 \pm A(r) \Gamma^{0}
\slash{f}\Gamma_m\epsilon^2 \ 
\end{array}
\eeqn
and
\beqn\label{eq:rindependent}
\begin{array}{rcl}
0&=& \pm\slash{F}\epsilon^2 + \half\slash\partial B \epsilon^1 \\
0&=&(D_m \pm \fourth H_m)\epsilon^1 + \slash{F} \Gamma_m \epsilon^2  \ .
\end{array}
\eeqn

Using (\ref{eq:rindependent}) and (\ref{eq:rdependent}), we can eliminate the 
R-R dependence in the $\delta \lambda = 0$ and $\delta \psi_r = 0$ equations.
The dilatino variation becomes
\begin{equation}
  \label{eq:dil}
0=(\slash{D} +\slash{\partial}(2B -\phi) )\epsilon^1 \pm\fourth
\slash{H}\epsilon^2 \ ; \qquad \ 
(\frac 14 g^{mn} g_{mn}' )\epsilon^1 =0 \ 
\end{equation}
while the radial gravitino equation becomes
\begin{equation}
  \label{eq:der}
0 = \partial_r\epsilon^1 -\frac12 U'\epsilon^1 -
\frac14 \slash b' \epsilon^1
\ .
\end{equation}
The final term in this equation is the only one which depends explicitly on 
the internal coordinates; we will take it to vanish separately from the other 
two terms. 

Similarly, one can eliminate the R-R dependence from the
$\delta\psi^1_{\theta,\phi}=0$ equations.  The result is
\begin{equation}
  \label{eq:deth}
  D_\alpha \epsilon^1 +\frac12 \gamma_\alpha \Gamma^1 \epsilon^1 = 0 \ , 
\qquad \alpha=(\theta,\phi)\ .
\end{equation}
where $D$ and $\gamma_\alpha$ denote the spin connection and gamma matrices
on a unit $S^2$.

We can now integrate (\ref{eq:der}) and (\ref{eq:deth})
to determine the spatial dependence of $\epsilon$. 
The radial equation implies that
\beqn
\epsilon^i(r,\theta,\phi,y) = e^{-\frac12 U(r)}\epsilon^i_0(\theta,\phi,y)
\,\ ,
\eeqn
where $\epsilon_0$ is independent of radius.
We will not need to write down the explicit dependence of $\epsilon_0$ on
$(\theta,\phi)$, but it is straightforward to do so using methods similar 
to those described in
\cite{lupoperahmfeld}. 

We can now decompose the radially independent, ten-dimensional
spinors $\epsilon_0^{1,2}$ in terms of four and six dimensional spinors, as 
\begin{equation}
\label{eq:10dspinoransatz}
\begin{array}{c}
\epsilon_0^1 = \zeta^1_+(\theta,\phi)\otimes\eta^1_+(y) +
\zeta^2_+(\theta,\phi)\otimes\eta^2_+(y) + \mathrm{c.~c.}
\\
\epsilon_0^2 = \zeta^1_+(\theta,\phi)\otimes\eta^3_{\mp}(y) + 
\zeta^2_+(\theta,\phi)\otimes\eta^4_{\mp}(y) +  \mathrm{c.~c.}
\end{array}
\end{equation}
Here $\zeta^i$ and $\eta^i$ denote four and six dimensional spinors,
respectively; a subscript on a spinor denotes its chirality.
The type IIA (IIB) case is given by the upper (lower) sign, where
$\epsilon^{1,2}$ have the opposite (same) chirality. 
Physically, $\zeta^{1,2}$ can be thought of as the two supercharges
in four dimensions that would be preserved if the black hole were
not present. 
The \cy\ case involves taking $\eta^2 =
\eta^3=0$ and $\eta^1=\eta^4$ to be the single globally defined spinor.

We can now insert (\ref{eq:10dspinoransatz}) into (\ref{eq:rdependent})
and collect terms of the same four-dimensional chirality.
These equations imply that
$\Gamma_0\zeta^2_- = \alpha(r)\zeta^1_+$, where
$\alpha(r)$ is an r-dependent phase that will be determined.
This relationship can be thought of as breaking the
$\calN=2$ supersymmetry that would have been preserved in four dimensions
down to a single linear combination.

Equation (\ref{eq:rdependent}) becomes 
\begin{equation}
  \label{eq:gravflow}
  \begin{array}{ccc}\vspace{.2cm}
-\frac12 U'\eta^1_+ \mp \alpha A(r) \slash{f} \eta^4_{\pm}= 0\ , &\qquad & 
\frac14 (-g\pm b)_{mn}'\gamma^n\eta^1_+ 
\mp\alpha A(r)\slash{f}\gamma_m \eta^4_{\pm}= 0\ , \\\vspace{.2cm}
-\frac12 U'\eta^2_+ \pm \alpha A(r) \slash{f} \eta^3_{\pm}= 0\ , &\qquad & 
\frac14 (-g\pm b)_{mn}'\gamma^n\eta^2_+ 
\pm\alpha A(r)\slash{f}\gamma_m \eta^3_{\pm}= 0\ , \\\vspace{.2cm}
-\frac12 U'\eta^4_+ - \alpha A(r) \slash f^\dagger 
\eta^1_{\pm}= 0\ , &\qquad & 
\frac14 (-g\pm b)_{mn}'\gamma^n\eta^4_+ 
-\alpha A(r)\slash f^\dagger\gamma_m \eta^1_{\pm}= 0\ , \\
-\frac12 U'\eta^3_+ + \alpha A(r) \slash f^\dagger 
\eta^2_{\pm}= 0\ , &\qquad & 
\frac14 (-g\pm b)_{mn}'\gamma^n\eta^3_+ 
+\alpha A(r)\slash f^\dagger\gamma_m \eta^2_{\pm}= 0\ . 
  \end{array}
\end{equation}

Equations (\ref{eq:rindependent}) and (\ref{eq:dil}) are precisely 
the equations that arise in the classification of Minkowski vacua 
in \cite{gmpt2}.  In the next section we will follow the analysis
of \cite{gmpt2} to study these equations.


\subsection{From variations to the attractor}
\label{sec:solve}
In this section we will analyze the spinor equations described above in terms
of geometrical quantities. In doing so, it will be
convenient to use 
the pure spinor formalism of \cite{hitchin,gualtieri,witt}, which is reviewed
briefly in appendix B.  We will 
focus on the IIB case, and use the lower sign in (\ref{eq:gravflow}).
The analysis for IIA is almost identical -- we will simply quote the 
IIA results at the end of this section.

We will start by reviewing the various structures defined by our spinors on
the internal manifold $Y$ 
-- this is described in greater detail in appendix B.  
In six dimensions, a single spinor $\eta$ 
with no zeros defines an SU(3) structure
on the tangent bundle $T$ of $Y$.  This is simply the statement that one can
form from this spinor two non-vanishing differential forms, 
a two form $J$ and three form $\Omega$, which obey 
$J\wedge \Omega=0$ and $J^3 = {3\over 4}i \Omega \wedge {\bar \Omega}$.  
These forms are associated to two elements $\slash{\Omega}$ and 
$\slash{e^{iJ}}$ of the Clifford algebra, which are given by exterior products 
of the original spinor: $\slash{\Omega} = \eta_+ \otimes \eta_-$ and 
$\sei = \eta_+ \otimes \eta_+$.  
  
For a pair of spinors, say $(\eta^1, \eta^3)$, 
the structure induced on the tangent bundle of $Y$ is more complicated -- 
it depends on the relative orientation of $\eta^1$ and $\eta^3$.
If the spinors are always parallel they define an SU(3) structure.  If they 
are orthogonal they define what is known as an SU(2) structure.  If they
are neither parallel nor orthogonal, they define what is sometimes known 
as a ``dynamic SU(2) structure.''

So far we have discussed the structures defined on the tangent bundle, but
it is more useful to discuss the structure defined on the sum of the 
tangent and cotangent bundles, $T\oplus T^*$. 
In fact, all of the cases described above define an SU(3)$\times$SU(3) 
structure on $T\oplus T^*$. 
To see this, first note that the bispinor
$\slash{\Phi}^{13}_{\pm}=\eta^1_+\otimes \eta^{3\dagger}_{\pm}$
defines a pair of SU(3) structures, 
via its annihilators from the left and from the
right. These two SU(3) structures live on $T\oplus T^*$, because
$\slash{\Phi}^{13}_{\pm}$ can be mapped via the Clifford map
to the bundle of differential forms, which is a representation of the Clifford
algebra on $T\oplus T^*$. 
These bispinors are known as pure spinors, because they are annihilated by
half of the elements of the algebra Clifford(6,6). 

When we have four spinors $\eta^1_+,\ldots, \eta^4_+$, the structures
are even more complicated. On the tangent bundle $T$, they can define 
anything from an SU(3) structure to a trivial structure (meaning
that $T$ is trivial and the manifold is parallelizable). 
The structure defined on the sum $T\oplus T^*$ 
can range from SU(3)$\times$SU(3) to SU(2)$\times$SU(2). The 
reason is that this time, even for the classification of vacua we need
more pure spinors: not just $\Phi^{13}_\pm$, but also $\Phi^{24}_\pm$, 
as we will see shortly.

We now turn to the analysis of the equations. 
We will start with  (\ref{eq:rindependent}) and (\ref{eq:dil}),
which are the same as the ones found in the classification Minkowski vacua.
In particular, 
the pairs $(\eta^1,\eta^3)$ and $(\eta^2, \eta^4)$ each {\it separately}
satisfy the equations for a Minkowski vacuum.  It is straightforward to
write these equations in terms of pure spinors.
In type IIB, one gets \cite{gmpt2} 
\begin{eqnarray}
  \label{eq:vacuaB+}
e^{-2B+\phi}(d-H\wedge)(e^{2B-\phi}\Phi^{13,24}_+)&=& 
dB\wedge \bar \Phi^{13,24}_+ + 
i\frac{e^\phi}8|\eta_{1,2}|^2 *\sigma(F) \ ,\\
\label{eq:vacuaB-}e^{-2B + \phi} (d- H\wedge) (e^{2B-\phi}\Phi^{13,24}_-) &=& 0 
\end{eqnarray}
where $\sigma(F_k)= (-)^{[k/2]}F_k$ as in \cite{gualtieri}. 
In addition, we also have $d\log|\eta_{1,2}|^2= dB$.  
We have used here the fact that, to have a supergravity
vacuum with fluxes, one needs an orientifold action; this relates
$\epsilon^1$ to $\epsilon^2$ in such a way that $|\eta^1|^2=|\eta^2|^2$ and
$|\eta^3|^2=|\eta^4|^2$ \cite{gmpt2}. In fact, we will see shortly
that all four of these norms are equal. 

To summarize, we have found that for any $r$ 
the internal manifold must support an $\nn=2$ vacuum.
In other words, 
the radial flow moves us through the moduli space of $\nn=2$ Minkowski vacua, 
just as we would expect.  

We now turn to the main computation of this paper: analyzing the equations
for radial evolution through the moduli space of $\nn=2$ vacua. 
We first look at the evolution of $U$. The first
equation in the first column of (\ref{eq:gravflow}) implies that
\begin{equation}
  \label{eq:Uprime}
  U'= 2\alpha A |\eta|^2 
\frac{(\sigma(f)\wedge \bar\Phi_-^{14})_{\mathrm{top}} }
{(\sigma(\Phi_-^{14})\wedge \bar\Phi_-^{14})_{\mathrm{top}} }=
\frac{e^U}{r^2}\left(
c\alpha(r) \frac{\int e^{2B+\phi}\sigma(f)\wedge \bar\Phi_-^{14}}
{4\int  \sigma(\Phi_-^{14})\wedge \bar\Phi_-^{14}}\right) \ .
\end{equation}
In the first step we have used $\mathrm{Tr}(\gamma\slash A \slash B)=
\frac{8 i}{\sqrt{g}} (A\wedge B)_{\mathrm{top}}$, and in the second step
we have used the fact that $U'$ is constant in the
internal directions. The appearance of $\sigma$ in this formula might seem 
unfamiliar; the pairing $(\sigma(A)\wedge B)_{\mathrm{top}}$
between differential forms $A$ and $B$, often denoted $(A,B)$, 
is known as the {\it Mukai pairing} (see for example
\cite{gualtieri}). The term in the parenthesis in (\ref{eq:Uprime}) 
may be thought of as the absolute value of the central charge of the
black hole, $|Z|$, which typically arises in the black hole attractor
equations -- a brief review of these equations is contained in Appendix C.
Note that the phase $\alpha$ is determined by (\ref{eq:Uprime}), since
$U$ is real. 

If we had instead used the third equation in the first
column of (\ref{eq:gravflow}), we would have obtained the same
equation with $|\eta_4|^2$ replaced by $|\eta_1|^2$.  This implies that 
$|\eta_4|^2= |\eta_1|^2$. We can derive similar 
equations for $\eta^2$ and $\eta^3$, from which it follows that all of the
spinor norms are equal.  These spinor norms are just given by 
$|\eta|^2= c e^{B}$ where $c$ is an integration constant 
(see the comment after (\ref{eq:vacuaB-})).
We have used this fact in writing (\ref{eq:Uprime}).

Since we have already factored the radial dependence out of $\epsilon_0$,
the bispinors $\slash{\Phi}^{14}_{\pm}$ are
independent of $r$: $\partial_r\slash{\Phi}^{14}_{\pm}=0$.
However, this does not mean that the differential forms $\Phi_\pm$ (related
to the bispinors by the Clifford map) are independent of radius. 
This is because the internal metric $g_{mn}$, and hence the gamma matrices 
$\gamma_m$, depend on the radius. 
For an odd bispinor $\slash{C}$,
\begin{equation}
  \label{eq:nog}
\partial_r(\slash{C})= 
\begin{picture}(10,10)(-15,5)
\put(0,0){\line(6,1){100}}
\end{picture}
(\partial_r C + \beta^m{}_n dx^n\wedge \iota_m C)
=\begin{picture}(10,10)(-15,0)
\put(0,0){\line(2,1){20}}
\end{picture}
(\partial_r C) + \frac12 \beta_{mn}\left( g^{mn}\slash C +\frac12 
\gamma^m \slash{C}\gamma^n\right)\ 
\end{equation}
where, as described in section 2.1, 
$\beta_{mn}= -\frac12 g'_{mn}$.
In the second step we have
used (\ref{eq:cl}). The resulting equation describes
the variation of an odd bispinor $\slash{C}$ 
due only to the variation of the components $C_{m_1\ldots m_k}$, 
after removing the contribution from the gamma matrices. 
The formula for even bispinors differs from (\ref{eq:nog}) by some signs.  

Let us consider the case where $\slash C=\slash\Phi^{14}_-=\eta_+^1\eta_-^{4\,\dagger}$.  
First, note that the $g'_{mn}g^{mn}$ term in (\ref{eq:nog}) vanishes 
due to (\ref{eq:dil}). 
We are left with a term of the form 
$g_{mn} \gamma^n \slash \Phi^{14}_- \gamma^m$, which by
(\ref{eq:gravflow}), can be written as the sum of
two terms, one proportional to $b'_{mn} \gamma^m \slash\Phi^{14}_-\gamma^n$ and
the other proportional to $\slash f \gamma_m \slash\Phi^{14}_- \gamma^m$. 
We will now describe how to massage these two terms. 

We will start with the $b'_{mn} \gamma^m \slash\Phi^{14}_- \gamma^n$ term.
The $m,n$ indices are antisymmetrized, so we can use the fact that
$\gamma^{[m} (\cdot) \gamma^{n]}=-dx^m\wedge dx^n \wedge+\iota^{mn}$.
Since $\slash b' \eta_\pm^i=0$, 
(see the discussion after (\ref{eq:der})), it follows that
$b'_{mn}[\gamma^{mn}, \slash \Phi_-^{14}]=0$; this can be rewritten as
$b'_{mn}(dx^m\wedge dx^n + \iota_{mn})\Phi^{14}_-=0$.  So we are just left with
$2 b'\wedge \Phi^{14}_-$. 

We can now attack the 
$\slash f \gamma_m \slash\Phi^{14}_- \gamma^m$ 
term, which is more interesting. The manipulation 
we will describe is similar to one used in \cite{gmpt,gmpt2}.\footnote{ 
From the description given in the text, it is not clear that we 
have extracted all of the information from (\ref{eq:gravflow}). 
To show that this is indeed the case, one can expand $f$ 
in terms of the pure Hodge diamond basis used in \cite{gualtieri,gmpt2}.
} 
Since $\slash\Phi_-^{14}=\eta_+^1\eta_-^{4\,\dagger}$, we can use
the first equation in the second column of (\ref{eq:gravflow}) 
to get $\slash f \gamma_m \eta_-^4 \eta_-^{4\,\dagger}\gamma^m$. 
We can then use the fact that $\frac{1+\gamma}2= 
\eta_+\eta_+^\dagger +\frac12 \gamma^m\eta_-\eta_-^\dagger \gamma^m $ for
any $\eta$; this is just the
expansion of the operator $\frac{1+\gamma}2$ in the chiral 
basis $\eta_+$, $\gamma^m \eta_-$. So the term under consideration can be 
written as a linear combination of $\slash f (1+\gamma)$ and 
$\slash f \eta^4_+ \eta_+^{4\,\dagger}$.  
Using the third equation in the first column of (\ref{eq:gravflow}), this 
second term is just $U' \slash\Phi_+^{14}$. 

One can similarly manipulate $\del_r \Phi^{14}_+$. The only 
difference
is that this time the $f$ contribution looks like $\slash f \gamma_m \eta^4_-
\eta^{4\,\dagger}_+\gamma^m$. This vanishes, because $\eta^4_-
\eta^{4\,\dagger}_+$ is (the slash of) a three--form and
$\gamma_m \slash C_k \gamma^m = (-)^k (6-2k)\slash C_k$.
 
If we put this all together, we 
get two equations involving bispinors $\slash{\Phi}^{14}_\pm$.
We can write these equations in terms of differential forms as
\begin{eqnarray}
  \label{eq:Phi+prime}
e^{b\wedge}\del_r\Big(e^{-b\wedge} \Phi^{14}_+\Big)&=&0\\ 
  \label{eq:Phi-prime}
e^{b\wedge}\del_r\Big(e^{-b\wedge} \Phi^{14}_-\Big)&=& 
 \alpha A |\eta|^2 ( f + i \sigma(*f)) - \alpha^2 U' \bar\Phi_-^{14} \ 
\end{eqnarray}
where again $\sigma(f_k)= (-)^{[k/2]}f_k$. 
It is interesting to note that these formulas are quite similar to
ones describing vacua.
This resemblance would be even more explicit if we had, e.g. considered a
non-compact $Y$ -- the norms of the spinors would not necessarily be equal, 
and we would have obtained an $F$ term in addition to $*F$. 
Finally, we should note that there is a similar pair
of equations for $\Phi^{23}_\pm$, which are found by taking 
${}^{14}\to {}^{23}$ and $\alpha\to -\alpha$. 
We will focus on only the $\Phi^{14}_\pm$ equations 
for the rest of this subsection.

These formulae, together with equation (\ref{eq:Uprime}) for $U'$, describe
how the geometry of the black hole and the internal manifold varies with radius. 
They are the generalizations of the 
attractor equations for this background, and one of the 
main results of this paper. Equation (\ref{eq:Phi+prime}) says that the four 
dimensional hypermultiplet moduli do not flow with radius.  
Equation (\ref{eq:Phi-prime}) describes how the four dimensional vector moduli flow. 
To compare these to the usual attractor equations 
it is useful to write them in a slightly different form.  
Taking the real part of (\ref{eq:Phi-prime}) gives
\begin{equation}
  \label{eq:Omega'}
\mathrm{Re}\, \frac{\del_r\Big(e^{-b\wedge} \Phi^{14}_-\Big)}
{A \alpha |\eta|^2} =
e^{-b\wedge} \Big[ f - 
\mathrm{Re}\left( \Phi^{14}_- \frac{U'}
{\alpha A |\eta|^2} \right)\Big]\ 
\end{equation}
which resembles the standard attractor flow equations. 

Near the horizon of the black hole at $r=0$ the geometry of $Y$
approaches an attractor fixed point.  At this fixed point, the 
pure spinor obeys a generalized stabilization equation
\beqn\label{eq:IIBattreqn}
f_1 + f_3 + f_5 = 2\mathrm{Im}( \Cbar\Phi^{14}_-) \ ,\qquad
\Cbar= \frac{i\int e^{2B+\phi} \sigma(f)\wedge \bar\Phi_-^{14}}
{e^{2B+\phi} \int  \sigma(\Phi_-^{14})\wedge \bar\Phi_-^{14}} \ .
\eeqn
Thus 
the charges of the black hole, in terms of the fluxes $f_1, f_3, f_5$, 
fix $\slash{\Phi}^{14}_-$ on the internal manifold.  
This is the generalization of the statement that, in type IIB, the holomorphic
three form of a Calabi-Yau is fixed by the charge of a BPS black hole.
We will demonstrate that this equation can indeed be solved 
in the following section, using a theorem of Hitchin's. 

We will simply quote the corresponding results for type IIA. 
The function $U'$ obeys
\begin{equation}
  \label{eq:UprimeA}
  U'= 
\frac{e^U}{r^2}\left(
c\alpha(r) \frac{\int e^{2B+\phi}\sigma(f)\wedge \bar\Phi_+^{14}}
{4\int  \sigma(\Phi_+^{14})\wedge \bar\Phi_+^{14}}\right) .
\end{equation}
The attractor equations obeyed by the pure spinors are
\begin{eqnarray}
  \label{eq:Phi-primeA}
e^{-b\wedge}\del_r\Big(e^{b\wedge} \Phi^{14}_-\Big)&=&0\\ 
  \label{eq:Phi+primeA}
e^{-b\wedge}\del_r\Big(e^{b\wedge} \Phi^{14}_+\Big)&=& 
 \alpha A |\eta|^2 ( f - i \sigma(*f)) + \alpha^2 U' \bar\Phi_+^{14} \ .
\end{eqnarray}
Again, from the four dimensional point of view this says that the 
vector multiplet moduli flow as a function of $r$.
Taking the real part of (\ref{eq:Phi+primeA}) gives
\begin{equation}
  \label{eq:eiJ'A}
\mathrm{Re}\, \frac{\del_r\Big(e^{b\wedge} \Phi^{14}_+\Big)}
{A \alpha |\eta|^2} =
e^{b\wedge} \Big[ f +
\mathrm{Re}\left( \Phi^{14}_+ \frac{U'}
{\alpha A |\eta|^2} \right)\Big]\ .
\end{equation}
At the attractor point this gives the stabilization equation
\beqn\label{eq:IIAattreqn}
f_0 + f_2 + f_4 + f_6 = 2\Im(\Cbar \Phi^{14}_+) \ , \qquad
\Cbar=\frac{i\int e^{2B+\phi} \sigma(f)\wedge \bar\Phi_+^{14}}
{e^{2B+\phi} \int  \sigma(\Phi_+^{14})\wedge \bar\Phi_+^{14}}  \ .
\eeqn
As in the IIB case, the constant can be determined from 
(\ref{eq:eiJ'A}) and (\ref{eq:UprimeA}), or by wedging both sides
with $\Phi_+$ and integrating. 

\subsection{Solving the attractor equation}
\label{sec:hitchin}
Equations (\ref{eq:IIBattreqn}) are a new version of the usual 
attractor equations, phrased in the language of pure spinors.
We can now use mathematical results concerning pure spinors, 
such as those of Hitchin
\cite{hitchin,hitchinsta,hitchin67} 
to describe the solutions to these equations.\footnote{
See \cite{glw} for a review of this mathematics in the 
context of four--dimensional effective theories.  
}  
These results determine 
exactly when a sum of differential forms can be the imaginary part 
of a pure spinor, in terms of a stability condition.  

Before discussing this theorem, let us make one comment about the 
attractor equation (\ref{eq:IIBattreqn}).  
First, note that the Bianchi identity (\ref{eq:bianchi})
implies that $\del_r(e^{-b\wedge} f)=0$.  So  
$f$ depends on $r$, as one would expect since the geometry of $Y$
changes as a function of radius. 
The value of $f$ at $r=\infty$ is related to the value at $r=0$ by
$f_\infty= f_{\mathrm{att}}e^{\Delta b}$, where
$\Delta b= b_\infty - b_{\mathrm{att}}$. 
The flux $f$ appearing in (\ref{eq:IIBattreqn}) is evaluated at 
the attractor fixed point, $f=f_{\mathrm{att}}$. 
So the attractor equation we are trying to solve is, when written in
terms of the flux at infinity,
$f_\infty= 2\Im ( \bar C e^{\Delta b} \Phi_-^{14})$. 

We are now in a position to state Hitchin's theorem; 
the ideas behind it are explained briefly
in appendix B and in the references. Given a sum of forms $f$, define
\begin{equation}
\label{eq:Hcriter}
q(f)=\mathrm{Tr}({\cal J}^2) \ , \quad 
{\cal J}_{\Lambda\Sigma}\equiv \frac{( \sigma(f)\wedge \Gamma_{\Lambda\Sigma} f)_{\mathrm{top}}}
{\mathrm{vol}}\ .
\end{equation}
Here $\Lambda$ and $\Sigma$ are indices on $T\oplus T^*$, 
as explained in the appendix. 
Then, $f$ is the imaginary part of a pure spinor $\Phi$ if and only if 
$q(f)<0$ everywhere (the quotient is understood pointwise). 
If this condition is satisfied, the pure spinor $\Phi$ 
is determined explicitly, as 
\begin{equation}
\label{eq:sol}
e^{\Delta b}
\bar C \Phi^{14}_-
= i\,f_\infty 
- \frac{{\cal J}_{\Lambda \Sigma}}{\sqrt{-q/12}} \Gamma^{\Lambda \Sigma} f_\infty
\ . 
\end{equation}
This is precisely the same pure spinor
$\slash \Phi^{14}_- = \eta_+^1 \eta_-^{4\,\dagger}$ that appeared
on the right hand side of (\ref{eq:IIBattreqn}).  
We should note that Hitchin's theorem describes a {\it pointwise}
obstruction to solving the attractor equation.\footnote{
However, we should note that the 
sign of $f\wedge \Phi\sim\bar\Phi\wedge\Phi$ at one point in the 
internal manifold determines the sign at every other point, since 
$|\eta|^2=ce^B$.}

The function $q$ described above is related to the Bekenstein-Hawking 
entropy of the black hole.  This entropy is given by the area of the horizon
in four dimensions, which depends on $U(r)$, and can be determined in 
terms of $f$ via (\ref{eq:Uprime}).  Plugging (\ref{eq:sol}) into
(\ref{eq:Uprime}) gives an expression for the entropy in terms
of the pure spinor $\Phi^{14}_-$ evaluated at the attractor fixed point:
it is essentially the square of the central charge $|Z|^2$, which is
$|\int \sigma(f) \wedge \Phi^{14}_-|^2$ times an appropriate normalization
factor.
In fact, this entropy can be written succinctly in terms of $q(f)$ as
\begin{equation}
S \sim \int e^{4B+2\phi} \sqrt{-q(f)} \ .
\end{equation}  
This relation between the entropy and $q(f)$ has been noted already by 
\cite{dgnv,pestun1}. 
We should emphasize that this construction gives a nice physical interpretation to 
Hitchin's  theorem:
one can solve the attractor equation precisely when the corresponding 
solution has positive (and real) entropy, i.e. when the black hole has a 
non-vanishing horizon.


There is one additional subtlety we have not yet discussed. 
The pure spinor $\Phi^{14}_-$ fixed by the attractor equations is implicitly 
related to the pure spinors describing the vacuum, $\Phi^{13,24}_\pm$, since
they are both built out of the same spinors.
In particular, equations (\ref{eq:vacuaB+}) and (\ref{eq:vacuaB-})
can be expressed as a rather complicated differential constraint on 
$\Phi^{14}_-$. 
We expect that 
this constraint can be solved by changing $f\to f+ dc$ for a suitable choice 
of $c$. This is the approach used in \cite{hitchin}, for the case where
$\Phi^{14}$ is closed. 
There, the existence of a suitable $c$ is reduced to a variational
problem for the integral of $q(f)$. 
This gives a moduli space of solutions as an open set 
in the appropriate de Rham cohomology.
In our case, this can be applied directly when $Y$ is a Calabi--Yau.
For example, when $Y$ is a torus case all spinors are covariantly constant and 
the differential constraints are trivial. More generally, 
if $\Phi^{14}$ is not closed, 
one would need to modify the Hitchin functional.  
We hope to be able to describe the general solutions of these constraints
in the future.

\section{Examples}

We will now consider a few particular cases of the general equations 
constructed above.  In section 3.1 we will describe how these equations 
reduce to the standard form in the Calabi-Yau case, before considering the 
explicit example of $T^6$ in section 3.2.  
We should emphasize that the examples considered in this section are  
meant to be illustrative of the techniques involved in solving the equations, 
but are probably not representative.

\subsection{The Calabi-Yau Case}

The four dimensional 
attractor equations in this case are well known; they are reviewed in 
appendix C. Here our approach differs from
existing ones only in that it is formulated in ten dimensions,
rather than in terms of a low energy $\nn=2$ supergravity in four dimensions.

When the internal manifold $Y$ is Calabi-Yau, it admits only one globally
defined spinor $\eta$. The ten-dimensional spinor ansatz is given by
(\ref{eq:10dspinoransatz}), with $\eta^2=\eta^3=0$ and
$\eta^1=\eta^4=\eta$. The pure spinors are related to the holomorphic
three form and Kahler form on $Y$, by
$\slash{\Phi}^{14}_- = \frac{i}{8}\slash{\Omega}$ and
$\slash{\Phi}^{14}_+ = \slash{e^{iJ}}$.  All other pure spinors vanish.

We will first consider the IIB case. 
The stabilization equation for the pure spinor at $r=0$, 
(\ref{eq:IIBattreqn}), becomes
\beqn
f_3 = 2 \Im \left( \Cbar\Omega \right),\ \ \ \ \ 
\Cbar =   \frac{i\int f_3\wedge \Omegabar }
{\int\Omega\wedge\Omegabar} \ .
\eeqn  
This is the usual stabilization equation (see, e.g. \cite{moore}).
In addition, one can verify that the radial flow in complex structure 
moduli space is precisely that described by (\ref{eq:IIBattreqn}).
The attractor equations at finite $r$ are typically written in terms of
a symplectic periods $(X^I, F_I)$ rather than directly 
in terms of the holomorphic three form $\Omega$.
For this reason, we have included in appendix C a discussion of the finite
$r$ attractor equation, formulated as a differential equation for 
$\Omega$.  It is straightforward to verify that this equation is just
(\ref{eq:IIBattreqn}).

The expressions in type IIA are identical, except that the two pure spinors
$\Phi^{14}_-$ and $\Phi^{14}_+$ are exchanged.  For example, 
the stabilization equation becomes
\beqn
f_0 + f_2 + f_4 + f_6 = 2\Im(\Cbar e^{iJ})
\eeqn
where the constant is fixed by
\beqn
\Cbar = \frac{i\int \sigma(f) \wedge e^{iJ}}{\int\sigma(e^{iJ})\wedge
e^{iJ} } = \frac{2\int (f_0 - f_2 + f_4 - f_6) \wedge
e^{iJ}}{\int J\wedge J \wedge J } \,\, .
\eeqn

\subsection{IIB on $T^6$}
\label{sec:t6}
Consider type IIB string theory 
compactified on $T^6$. For most of this section 
we will not consider orientifolds of $T^6$; they will be
discussed briefly at the end of the section.
Without orientifolds or flux, type IIB on $T^6$ gives an
$\calN=8$ supergravity in $d=4$. 
The field content is a single $\calN=8$ gravity
multiplet, which contains 70 real scalar fields and 28 vector
fields.  There are 56 objects charged under these gauge fields:
\begin{center}
\begin{tabular}{cccc}
gauge\, field &electric\, object & magnetic\, object & number \\
$C_{\mu abc}$  & D3 & D3 & 20 \\
$C_{\mu a}$ & D1 & D5 & 12 \\
$B_{\mu a}$ & F1 & NS5 & 12 \\
$g_{\mu a}$ & KK\, momentum & KK\, monopole & 12 \\ 
\end{tabular}
\end{center}

The attractor mechanism for black holes made out of D3 branes 
on the torus is a special case
of the usual Calabi-Yau attractor equations, and is described nicely in
\cite{moore}. 
Using the pure spinor formulation, it is straightforward to write down
analogous attractor equations for black holes made out of D1 and D5 branes
as well.  This provides a simple illustration of the power of pure spinor
techniques.

First, we must decide the form of the spinor ansatz 
(\ref{eq:10dspinoransatz}).  There are many possibilities.  
The simplest is to consider the torus as a \cy, which 
means taking $\eta^2=\eta^3=0$.  We will use this ansatz in what follows, 
because it is the simplest: in general, $\eta^2$ and $\eta^3$ 
will not be zero, and one will have
to solve the extra equations for the resulting pure spinor $\Phi^{23}$. 

We will now describe the pure spinor 
$\slash\Phi^{14}_-=\eta^1_+\eta^{4\dagger}_-$ on $T^6$.
If $\eta^1$ and $\eta^4$ are equal (as in
the Calabi-Yau case, where there is only one globally defined spinor),
the one form and five form pieces of $\Phi^{14}_-$ vanish.  This can
be seen easily by using a basis where the $\gamma_m$ are antisymmetric. 
The attractor equations in this case reduce to those described by \cite{moore}.

In general, however, $\eta^4$ will not be proportional to $\eta^1$.
The pure spinor $\eta^1_+ \otimes \eta^{4\dagger}_-$ will 
be of the form \cite{jw}
\beqn
\Phi^{14}_- =  \Omega +  e^{ij}\wedge v\ .
\eeqn
The first term is due to the component of $\eta^4$ parallel to 
$\eta^1$, and the second term is due to the component perpendicular to $\eta^1$. 
So the attractor equation for a configuration of D1, D3 and D5 branes on a torus is
\beqn
f_1 + f_3 + f_5 = 2\Im\left(\Cbar ( \Omega +  e^{ij}\wedge
v)\right)\ .
\label{eq:atttorus}
\eeqn
As we saw in section \ref{sec:hitchin}, 
we can determine whether this equation has a solution by looking at the charges. 
However, in order to illustrate the existence of new solutions, 
we can proceed in the opposite direction; first we choose 
a pure spinor, and take $f$ to be its imaginary part.
For example, we may choose the pure spinor to be   
$dz^1 (dz^2 dz^3+e^{(1/2i)(dz^2\bar dz^2+dz^3\bar dz^3})$. 
This leads to the charges $f_1=dx^1$, 
$f_3= dx^1dx^2 dx^3- dy^1dy^2 dx^3-
dy^1 dx^2 dy^3 - dx^1 dy^2 dy^3 + dx^1dx^2 dy^2 +dx^1 dx^3 dy^3$, and
$f_5=dx^1 dx^2 dy^2 dx^3 dy^3$.
This choice leads to a square torus (with all $\tau=i$)
and a finite value of the black hole area and entropy.

Finally, we can discuss more complicated cases, where the $T^6$ is 
orientifolded.  Consider the orientifold that reverses all the coordinates
on $T^6$, which generates $2^6$ O3 planes. 
This projection leaves invariant only the $C_{a\mu}$ and $B_{a\mu}$ gauge 
bosons. We must now choose a spinor ansatz that is
compatible with the orientifold action. 
This constrains $\eta^1_+=i\eta^3_+$ and $\eta^2_+=i\eta^4_+$, so that
$\Phi^{23}_-= \sigma(\Phi^{14}_-)$.  This is compatible with the fact that 
only $f_1$ and $f_5$ charges are allowed, since 
the three--form parts of $\Phi^{14,23}_-$ vanish.
One can easily prove in this case that no such pure spinors exist. 
This can be seen by noting that if a pure spinor
starts with a one form $v$, 
it will necessarily be of the form $v e^{\mathrm{form}_2}$, by a theorem 
in \cite{gualtieri}.  It can also 
be proven using the theorem in section \ref{sec:hitchin}. 
(Remember that there are no differential constraints in this case, since the 
spinors are all covariantly constant.) If 
we call $\tilde f_5$ the {\it vector} dual to the form $f_5$ 
(so that we do not need to use the metric),  then 
$q(f_1+f_5)= 6(f_1\tilde f_5)^2\geq 0$. 
This shows that these charges lead to no solution.

The previous discussion assumed that the charge of the orientifold is balanced
by D3 branes. One could ask what would happen if there are
$H\wedge F_3$ terms as well -- this is perhaps the simplest example
of a flux compactification. 
In addition to the problem described above, an additional constraint arises 
from the Bianchi identities. In particular, the $H$ flux generates new terms  
of the form $H\wedge f_1$ and $H\wedge *f_5$. Canceling both
would require either taking $H=0$ (as we did above) or $f_1=f_5=0$.

From the four--dimensional point of view, the vectors coming from the R-R
sector alone are not enough to have a non--singular solution in the orientifold case. 
One would have to 
mix the charges with those coming from the NS-NS sector. It would be 
interesting to extend the present work to incorporate these charges.\footnote{
For example, we may consider 
a solution with NS-NS charge $H_3 \sim \vol_A \wedge h^A + \vol_S \wedge h^S$.
This leads to stabilization equation $0 = \Im(C (\iota_h \Phi_+
+ h\wedge \Phi_+))$ where $h=h^A + i h^S$.}
Similar considerations apply to the simple non--\ka\ vacua introduced
in \cite{kst} by acting with T--duality on the torus with $F_3$ and $H$. 
Indeed, we would expect that one could write down simple attractor equations
in these cases, as the $\nn=2$ examples constructed in \cite{kst} are dual
to $\nn=2$ \cy\ compactifications \cite{Schulz:2004tt}.

Finally, we should mention that many of the considerations in this subsection
apply easily to $K3\times T^2$. This is $\nn=4$ rather than $\nn=8$, so 
the choice of internal spinors is more limited. In this
case we obtain an attractor equation of the form 
(\ref{eq:atttorus}), where $\omega$ and $j$ are
members of the triplet of covariantly constant two--forms on  $K3$.

{\bf Acknowledgments}
We would like to thank P.~Aspinwall, B.~Florea, F.~Denef, M. Gra\~ na, 
S.~Kachru, R.~Kallosh, X.~Liu, M.~Schulz, A.~Simons and A.~Strominger
for helpful discussions. J.H. is supported by NSF grant 0244728. 
The work of A. M. is
supported by the Department of Energy, under contract DE--AC02--76SF00515.
A.~T.~is supported by the DOE 
under contract DEAC03-76SF00515 and by the NSF under contract 9870115.

\appendix

\section{Gamma matrix conventions}

Our conventions for the four and six dimensional gamma matrices are
\begin{equation}
\begin{array}{rclcrcl}
\gamma_0 & = & \gamma_0^* = - \gamma_0^{\dagger}  \\
\gamma_i & = & \gamma_i^* = \gamma_i^{\dagger}  \\
\gamma_m & = & -\gamma_m^* =  \gamma_m^{\dagger}  \\
\end{array}
\end{equation}
where the four dimensional indices $\mu=(0,i)$ are raised and lowered with the flat
Lorentzian metric and the six dimensional indices $m$ are raised and lowered
with the flat Euclidean metric. With these definitions, the chirality
matrices are
\begin{equation}
\begin{array}{ccccccc}
\gamma_5 & = & i\gamma_0 \ldots \gamma_3 & = & - \gamma_5^* & = &
\gamma_5^{\dagger} \\
\gamma_7 & = & -i\gamma_4 \ldots \gamma_9 & = & - \gamma_7^* & = &
\gamma_7^{\dagger} \\ .
\end{array}
\end{equation}
Note that with our conventions $\gamma_5$ is pure imaginary.
The ten dimensional gamma matrices are
\begin{equation}
\Gamma_{\mu} = \gamma_{\mu} \otimes 1 , \qquad
\Gamma_m=\gamma_5\otimes\gamma_m
\end{equation}
and the ten dimensional chirality matrices are
\begin{equation}
\Gamma_5=\gamma_5\otimes 1 \qquad \Gamma_7 = 1\otimes\gamma_7 \qquad
\Gamma_{11} = \gamma_5 \otimes \gamma_7 = \Gamma_{11}^{\dagger} .
\end{equation}
\section{Pure spinors}
\label{jazz}

The objects $\Phi$ and $\slash\Phi$ 
considered in this paper have geometrical significance. The first lives in the
bundle of differential forms, and the second in the space of bispinors. 
The two are related by the Clifford map, which sends a form $dx^{m_1\ldots
m_k}$ to $\gamma^{m_1\ldots m_k}$. In this paper we
 denote the bispinor corresponding
to a differential form $C$ by $\slash C$. 

The space of bispinors can be viewed as the representation space for two
Clifford(6) algebras, which act by left and right multiplication.  The
space of differential forms can be viewed as the representation space
for an algebra generated by wedging with one forms, $dx^m\wedge$, and 
contracting with vectors, $\iota_m\equiv \iota_{\del_m}$. 
This algebra is call Clifford(6,6).  It is generated by {\it twelve} 
gamma matrices (identified with $dx^m\wedge$ and $\iota_m$) and has an
indefinite metric (given by the pairing between vectors and one--forms). 
It is sometimes useful to denote these twelve gamma matrices
collectively as $\Gamma^\Lambda$.

The action of Clifford(6)$\times$Clifford(6) on the space of bispinors is
related to the action of Clifford(6,6) of differential forms.  For an even (odd)
differential form $C_\pm$,
\begin{equation}
  \label{eq:cl}
\gamma^m \ \slash{C_\pm} = 
\begin{picture}(10,10)(-10,5)
\put(0,0){\line(6,1){100}}
\end{picture}
[(dx^m\wedge + g^{mn}\iota_n)C_\pm]
\ , \qquad 
 \slash{C_\pm}\ \gamma^m = \pm  
\begin{picture}(10,10)(-10,5)
\put(0,0){\line(6,1){100}}
\end{picture}
[(dx^m\wedge - g^{mn}\iota_n)C_\pm] \ .
\end{equation}
The observation that this Clifford product is represented by 
a combination of wedging and contracting is an old one, 
see e.g. \cite{lm}.  This relation 
is also apparent in the identities used to manipulate
products of antisymmetrized gamma matrices $\gamma^{m_1\ldots m_k}$, as in 
e.g. the appendix of \cite{cr}. Recently, this fact has been used in the
context of generalized complex geometry \cite{gmpt,witt,jw,gmpt2}.

A pure spinor $\Phi$ 
is annihilated by a dimension six subspace of Clifford(6,6) -- 
i.e. by six linear combinations
of $dx^m\wedge$ and $\iota_m$. To see why this definition is useful, consider
the sum of the tangent and cotangent bundles  $T\oplus T^*$. 
In general, the structure group of this bundle is O(6,6). However, 
the existence of a pure spinor $\Phi$ allows us to restrict this structure group
to the subgroup of $O(6,6)$ that leaves $\Phi$ invariant.
This turns out to reduce the structure group to SU(3,3). 

To see how this works, consider the space of annihilators of $\Phi$, 
which has dimension six.  This space can be viewed as the $(1,0)$
space of an almost complex structure ${\cal J}$ on $T\oplus T^*$,
which restricts the structure group to U(3,3). 
This ${\cal J}$ is known as a generalized (almost) 
complex structure because it lives on $T\oplus T^*$ rather than $T$.
This complex structure can be computed explicitly -- it is the 
expression given in equation (\ref{eq:Hcriter}).  
To understand the origin of this formula, 
remember that an ordinary almost complex structure
can be defined from an ordinary spinor $\eta_+$ as $iJ_{mn}=\eta_+ \gamma_{mn} \eta_+=
\Re(\eta_+) \gamma_{mn} \Re(\eta_+)$. 
The expression in (\ref{eq:Hcriter}) is the same, but with
$T\oplus T^*$ indices. 
It is now clear why Hitchin's criterion is necessary: since 
${\cal J}$ is an almost complex structure, it must obey 
(with appropriate normalization) $\mathrm{Tr}({\cal J}^2)= -12$.\footnote{
For spinors on $T$ we usually do not discuss criteria of this form. 
This is because in six dimensions all Clifford(6) spinors are pure.}

We can now ask what happens if the geometry 
admits a {\it pair} of pure spinors $\Phi_\pm$.
This pair allows us to 
reduce the structure group on $T\oplus T^*$ to $SU(3)\times SU(3)$. 
For the geometries described in this paper, these
pairs appear as exterior products of ordinary spinors, of the form
 $\Phi^{13}_\pm= \eta^1_+ \eta^3_\pm$.  Of course, not all pairs of pure spinors
can be written in this product form; they must obey a compatibility condition.
This compatibility condition implies, for example, 
that the intersection of the two spaces of annihilators has dimension 3. 
A pair of pure spinors $\Phi_\pm$ can be used to define a metric \cite{gualtieri}
\begin{equation}
\label{metric} g^{mn}= {\cal J}_+^m{}_p {\cal J}_-^{pn} + {\cal J}_+^{mp} {\cal J}_{-\,p}^n\  , \quad
{\cal J}_{\pm\,\Lambda\Sigma}
\equiv \frac{( \sigma(\mathrm{Re}(\Phi_\pm))\wedge \Gamma_{\Lambda\Sigma} \Re(\Phi_\pm))_{\mathrm{top}}} {\mathrm{vol}} \ .
\end{equation}
Given this metric, 
one can use (\ref{eq:cl}) to map the annihilators of the $\Phi_\pm$
to a subspace of dimension 3 of the left Clifford(6) action -- this subspace is
precisely the annihilator of the left $\eta^1_+$.   Likewise, 
once can also construct the annihilators of the right $\eta^3_\pm$.
These spaces of left and right annihilators define
an SU(3)$\times$SU(3) structure on $T\oplus T^*$. 

In the main text, the vacua under consideration are 
characterized by {\it two} pairs of pure spinors, $\Phi^{13}_\pm$ 
and $\Phi^{24}_\pm$.  Of course, these two pairs are not independent: they must
define the same metric. 

We should also mention an important special case, where the two spinors that define
$\slash\Phi_\pm = \eta_+\eta_\pm$ are equal.  
In this case we obtain an SU(3) structure on $T$. In general, one can compute
the explicit form of $\slash\Phi_\pm$ using Fierz identities. For the case at hand, 
it turns out that $\slash\Phi_-\equiv\frac i8 \slash{\Omega}$ for 
a complex three--form $\Omega$, and
$\slash\Phi_+= \frac 18 \sei $ for a real two--form $J$. 
In this case, the compatibility condition between the pure spinors is that
$J\wedge \Omega=0$ (i.e. $J$ is a $(1,1)$ form), 
and that 
$J^3=\frac34 i\Omega\bar\Omega$.
Together, $J$ and $\Omega$ provide an equivalent way of characterizing
the SU(3) structure of $T$.
They also define a positive signature metric,
 $g_{i\bar j}= i J_{i\bar j}$.

Finally, we should mention \cy\ case.
Here, the spinor is covariantly constant ($D_m \eta=0$) and the differential
forms are closed ($dJ=0=d\Omega$). These two conditions are equivalent.



\section{The four dimensional attractor equations}

In this appendix we will review the four dimensional attractor 
equations (see also \cite{moorelectures} for a review), 
and demonstrate that they are 
equivalent to the form described in the text.  

For a \cy\ compactification of type II supergravity, the low energy theory is 
D=4, $\nn=2$ supergravity with some number of vector and hyper multiplets.
The low energy theory includes BPS black hole solutions, 
whose metric is of the form
\begin{equation}
ds^2 = - e^{2U(r)} dt^2 + e^{-2U(r)}(dr^2 + r^2 (d\theta^2 + \cos^2 \theta d\phi^2) \ .
\end{equation}
The metric factor 
$U(r)$ and the vector multiplet moduli $t^a(r)$ are functions of radius.  
The BPS equations for this background reduce to a set of linear 
differential equations for $U$ and $t^a$, 
\beqn
{\dot U} = {e^U \over r^2} |Z|,\quad
{\dot t}^a = {e^U\over r^2 |Z|} g^{a\bar b} \p_{\bar b} |Z|^2 . 
\eeqn
Here $\cdot$ denotes $d/dr$, 
$Z(t^a(r))$ is the central charge and $g_{a\bar b}$ is the metric on
vector multiplet moduli space.  
For the rest of this appendix, we will describe using geometric language
the attractor equations in type IIB. 

For type IIB on a Calabi-Yau the moduli $t^a$ describe deformations of the complex
structure, which is related to the holomorphic three form $\Omega$.
The charge of the black hole is parameterized by an element $F$ of $H^3$.
The metric is K\"ahler,
\beqn
g_{a\bar b} = K_{a\bar b},\quad e^{-K}= i\prod{\O}{\bO}
\eeqn
and the central charge is
\beqn
Z = e^{K/2} \prod{\O}{F} \ .
\eeqn
The subscripts ${}_{a,\bar b}$ denotes derivatives, and we have defined
$\prod{\alpha}{\beta} = \int_Y \alpha \wedge \beta$.  
Since we are considering only the four dimensional effective theory, 
we may regard a three form on $Y$
not as a full differential three form but rather as 
(the harmonic representative of) an element of $H^3$.
So $\prod{~}{~}$ may be thought of as the symplectic inner product on a 
finite dimensional vector space.


Now, since $\O$ and $\bO$ are basis elements of $H^{3,0}$ and $H^{0,3}$, it 
is useful to define projection operators
\beqn
P^{3,0} \a = {\prod{\a}{\bO}\over\prod{\O}{\bO}} \O
,\quad
P^{0,3} \a = {\prod{\O}{\a}\over\prod{\O}{\bO}} \bO
.\eeqn
We will denote projection operators onto the transverse space 
by $P^{3,0}_\bot=1-P^{3,0}$ and $P^{0,3}_\bot=1-P^{0,3}$.
The derivatives $\O_a$ and $\bO_\ba$ are in $H^{3,0}\oplus H^{2,1}$
and $H^{0,3}\oplus H^{1,2}$, respectively, so we 
can define projection operators onto $H^{2,1}$ and 
$H^{1,2}$ by
\beqn
P^{2,1} \a = 
-g^{a\bb}{\prod{\a}{P^{0,3}_\bot\bO_\bb}\over\prod{\O}{\bO}} P^{3,0}_\bot \O_a
,\quad
P^{1,2} \a = 
-g^{a\bb}{\prod{P^{3,0}_\bot\O_a}{\a}\over\prod{\O}{\bO}} P^{0,3}_\bot \bO_\bb
.\eeqn
It is straightforward to verify that 
all of these projection operators obey $P^2=P$, commute with one another,
and are adjoints with respect to the symplectic inner product:
\beqn
\prod{\a}{P^{0,3} \beta} = \prod{P^{3,0} \a}{\beta} 
,\quad
\prod{\a}{P^{1,2} \beta} = \prod{P^{2,1} \a}{\beta} 
.
\eeqn
To show this, it is useful to use the explicit form of the metric
\beqn
g_{a \bar b}
= -{\prod{P^{3,0}_\bot \O_a}{\bO_\bb}\over \prod{\O}{\bO}} 
= -{\prod{\O_a}{P^{0,3}_\bot\bO_\bb}\over \prod{\O}{\bO}} .
\eeqn
Moreover, the operators described above form a complete basis of $H^3$, so
\beqn
P^{3,0} + P^{2,1} + P^{1,2} + P^{0,3} = 1.
\eeqn

It is straightforward to show that 
\beqn
\partial_a|Z|^2 = i {\prod{F}{\bO} \over \prod{\O}{\bO}}\prod{P^{3,0}_\bot\O_a}{F}
,\quad
\partial_\ba|Z|^2 =  i {\prod{\O}{F} \over \prod{\O}{\bO}}\prod{F}{P^{0,3}_\bot \bO_\ba} .
\eeqn
Using the fact that $\dot \O= {\dot t}^a \O_a$,
we can multiply both sides of the second attractor equation by $g_{a \bb}$ to get 
\beqn
\prod{\dot \O +i {e^U\over r^2 }{\prod{\O}{F}\over|Z|} F}
{P^{0,3}_\bot\bO_\ba}=0
,\quad
\prod{P^{3,0}_\bot\O_a}
{\dot \bO +i {e^U\over r^2 }{\prod{F}{\bO}\over|Z|} F} =0
.\eeqn
These equations fix the components of
$\dot \O$ and $\dot \bO$ in $H^{2,1}$ and $H^{1,2}$, respectively, so that
\beqn \label{eq:ProjectedOmegaDot}
P^{2,1}{\dot \O} = 
-i {e^U\over r^2} {\prod{\O}{F}\over |Z|} P^{2,1}F
, \quad
P^{1,2}{\dot {\bar\O}} = 
-i {e^U\over r^2} {\prod{F}{\bO}\over |Z|} P^{1,2}F.
\eeqn 

The $H^{3,0}$ and $H^{0,3}$ components are left unfixed, so we can write
\beqn
{\dot \O} = -i {e^U\over r^2} {\prod{\O}{F}\over |Z|} P^{2,1} F + \chi \O
,\quad
{\dot \bO} = -i {e^U\over r^2} {\prod{F}{\bO}\over |Z|}P^{1,2}F + \bchi \bO
\eeqn
for an arbitrary function $\chi(r)$.  
These are unphysical components of $\dot \O$ which 
can be absorbed into K\"ahler transformations on moduli space.  
Recall that the moduli space metric, and indeed the entire low energy action, 
are invariant under the K\"ahler transformation
\beqn
\O\to e^{f(t^a)}\O,\quad
\bO\to e^{\barf(\bt^\ba)} \bO
\eeqn
for an arbitrary holomorphic function $f(t^a)$ on moduli space.  
This transformation
takes $K \to K - (f(t^a) + \barf(\bt^\ba))$ and 
\beqn
\chi \to \chi + {\dot f} = \chi + f_a {\dot t}^a .
\eeqn
So, by judicious choice of $f(t^a)$, we may
set $\chi(r)$ to be whatever we like.  

One natural choice is
$\chi=0$.  In this case $\prod{\dot \O}{\bO} = \prod{\O}{\dot\bO}=0$, and 
the K\"ahler potential is independent of radius.  
Another simple choice is to take
\beqn
{\dot \O} = -i {e^U\over r^2} {\prod{\O}{F}\over |Z|} (P^{3,0} + P^{2,1})F
,\quad
{\dot \bO} = -i {e^U\over r^2} {\prod{F}{\bO}\over |Z|} (P^{0,3} + P^{1,2})F
.\eeqn
In this case
\beqn
\prod{\dot \O}{\bO} = \prod{\O}{\dot\bO} = i {e^U\over r^2} |Z|e^{-K}
\eeqn
is pure imaginary, so
\beqn
{\dot K} = 2 {e^U |Z| \over r^2} = 2 \dot U.
\eeqn
This equation can be integrated to give the relation of \cite{att1} 
between the spacetime and moduli space metrics:
$2 U(r) = K(r) - K(\infty)$.

However, it is useful to have a more explicit form for the attractor equations 
that does not require fixing K\"ahler gauge invariance.
First, note that 
\beqn
\dot K = - (\chi + \bchi)
,\quad
\p_r \left(\ln {Z\over \bZ}\right) = \chi-\bchi .
\eeqn
So $\chi = - {\dot \Theta} / \Theta$, where
\beqn
\Theta = \sqrt{{\bZ\over Z}}e^{K/2}
.\eeqn
With a little algebra, the attractor equations become
\beqn
\p_r (\Theta \O) = -i {e^U \over r^2 } P^{2,1} F
,\quad
\p_r (\bar{\Theta} \bO) = i {e^U \over r^2 } P^{1,2} F
.
\eeqn
Using the completeness relation for projection operators, these 
are equivalent to the single equation
\beqn\label{4dcyflow}
\Im\left\{\p_r\left( \Theta \O\right)\right\} = 
-{e^U\over 2r^2} \left(F-\Im 2 e^{K/2}\Zbar\Omega \right) 
\eeqn
where $\Theta$ is defined above.  Note that under K\"ahler 
transformations $\Theta\to e^{-f} \Theta$, 
so both sides of this equation are invariant.
It is straightforward to show that the equations in this form are equivalent to the 
pure spinor equation (\ref{eq:IIBattreqn}) described in the text.


\end{document}